\documentstyle[11pt,newpasp,twoside,psfig]{article}
\begin{document}

\title{Molecular Gas in nearby Early-Type Powerful Classical Radio Galaxies}
\author{St\'ephane Leon}
\affil{Instituto de Astrof\'{\i}sica de Andaluc\'{\i}a, Camino Bajo de Hu\'etor, 24
Apdo 3004, 18080 Granada, Spain}
\author{Jeremy Lim}
\affil{Institute of Astronomy \& Astrophysics, Academia Sinica, P.O. Box 23-141,
Taipei 106, Taiwan}
\author{Fran\c{c}oise combes}
\affil{Observatoire de Paris, LERMA, 61 Avenue de l'Observatoire, 75014 Paris, France}
\author{Dihn-V-Trung}
\affil{Institute of Astronomy \& Astrophysics, Academia Sinica, P.O. Box 23-141,
Taipei 106, Taiwan}

\setcounter{page}{111}
\index{Leon, S.}
\index{Lim, J.}
\index{Combes, F.}
\index{Dihn-V-Trung}

\begin{abstract}
We report a survey for molecular gas in nearby powerful radio galaxies. Eight of the 
eighteen radio galaxies observed were detected with molecular masses in the range 
$10^7$--$10^9 {\rm \ M_{\odot}}$, similar to the same survey we performed towards 3C radio
galaxies. The upper limits of molecular gas in the 
remainder are typically of $10^8 {\rm \ M_{\odot}}$, indicating that very few radiogalaxies 
have molecular gas reservoir with more than 10$^9 {\rm \ M_{\odot}}$.
\end{abstract}

\section{Introduction}
Powerful classical double-lobed  radio galaxies are (at low 
redshifts) hosted almost exclusively by luminous early-type galaxies.  These galaxies are usually 
elliptical galaxies. 
Given that normal elliptical galaxies usually exhibit undetectable quantities of cool (neutral) gas,
 what fuels the central supermassive black holes in elliptical radio galaxies? 
Our $^{12}$CO survey of a complete sample of 3C galaxies at low redshift (Lim et 
al., this volume) reveals that the vast majority have 5$\sigma$  upper limits in molecular gas 
masses of 2-5 x 10$^8 {\rm \  M_{\odot}}$ (for a linewidth of ~500 km/s and a Galactic 
CO-to-H$_2$ conversion factor).  To better 
understand the molecular gas content of classical radio galaxies, we are undertaking a survey of 
nearby early-type powerful radio galaxies.  These galaxies were originally selected for study by 
Verdoes Kleijn et al. (1999) with the HST, and were drawn from a catalog of one-hundred seventy-six 
radio-loud galaxies constructed by Condon \& Broderick (1988). 



\section{Results}
Of the eighteen galaxies where sensitive 12CO measurements have now been made, we detected eight 
(i.e., a detection rate of 45\%).  The  galaxies have molecular gas masses in the 
range 10$^7$-- 10$^9 {\rm \ M_\odot}$.  All 
the galaxies  detected have double-horned 12CO line profiles characteristic of a rotating disk 
or torus, commensurate with the disk-like dust features 
seen in these galaxies.  In the case of 3C 31, our follow-up imaging observations with the IRAM 
PdBI confirm that the molecular gas is distributed in a rotating disk/torus with size similar 
to its dust disk.  
These results, summarized in Table 1, reinforce our argument (Lim et al. 2000) that, in the 
vast majority of cases, 
the molecular gas in powerful classical radio galaxies most likely originates from a merger 
between a pre-existing elliptical galaxy and a relatively less massive gas-rich (disk) galaxy.
 Radio activity is triggered only when sufficient gas reaches the nucleus of the elliptical galaxy, 
thereby explaining the relatively relaxed nature of both the accreted molecular gas and parent 
elliptical galaxy.

\begin{table}[h]
\caption{Results of CO survey of nearby powerful  radio galaxies}
\begin{tabular}{lcccc}
\tableline
\tableline
Name     &  Redshift  &  IRAS  &  Optical Dust  &         $M{\rm(H_2)}$          \\
         &    (z)     &          &                &       $(\rm M_{\odot})$        \\
\tableline
NGC 193	 &   0.014    &		 & $\surd$	  &   $<1.2 \times 10^8 $    \\
NGC 315  &   0.016    &	$\surd$  & $\surd$ 	  &  $(3.03 \pm 0.34) \times 10^8 $ \\
NGC 383 (3C 31) & 0.017	 & $\surd$ & $\surd$	  &  $(1.06 \pm 0.04) \times 10^9 $ \\
NGC 541  &	0.018	& 	& $\surd$		&   $(2.47 \pm 0.38) \times 10^8 $ \\
UGC 1841 (3C 66B) & 0.021  & 	& 		& $ < 2.8\times 10^8 $ \\
NGC 2329 & 0.019	& $\surd$   &		& $ < 1.4 \times 10^8 $	\\
NGC 2892 & 0.023  &	&  		& $ < 3.4 \times 10 $	\\
NGC 3801 & 0.011	&      & $\surd$    & $ (3.77 \pm 0.22) \times 10^8 $	\\
NGC 3862 (3C 264) & 0.022  & $\surd$ & $\surd$ & $ (3.09 \pm 0.26) \times 10^8 $ \\
NGC 4261 (3C 270) & 0.007  & $\surd$ & $\surd$  &  $< 3.6 \times 10^7 $	\\
NGC 4335    &  0.015  &     & $\surd$   &  $  < 1.2 \times 10^8 $	\\
NGC 4374 (3C 272.1) & 0.003 & $\surd$ & $\surd$ & $ (1.40 \pm 0.38) \times 10^7 $ \\
NGC 4486 (3C 274)  & 0.007  & $\surd$  & 	  &  standing wave ripples \\
NGC 5127 & 0.016   &        & $\surd$  &   $ <  1.4 \times 10^8 $	\\ 
NGC 5141 & 0.018	& 	& $\surd $	&  $ <  4.4 \times 10^8 $ \\
NGC 7052 & 0.016   &  $\surd$    & 	&  $(3.45 \pm 0.23) \times 10^8 $ \\
UGC 12064 (3C 449) & 0.018 & tentative &  $\surd$ & $(2.85 \pm 0.23) \times 10^8 $ \\
NGC 7626   & 0.011 &    & $\surd$  &   $ <  0.8 \times 10^8 $	\\ 
\tableline
\end{tabular}
\end{table}

\end{document}